# Event Cause Analysis in Distribution Networks using Synchro Waveform Measurements

Iman Niazazari, Hanif Livani, Amir Ghasemkhani, Yunchuan Liu, and Lei Yang

*Abstract*—**This paper presents a machine learning method for event cause analysis to enhance situational awareness in distribution networks. The data streams are captured using time-synchronized high sampling rates synchro waveform measurement units (SWMU). The proposed method is formulated based on a machine learning method, the convolutional neural network (CNN). This method is capable of capturing the spatiotemporal feature of the measurements effectively and perform the event cause analysis. Several events are considered in this paper to encompass a range of possible events in real distribution networks, including capacitor bank switching, transformer energization, fault, and high impedance fault (HIF). The dataset for our study is generated using the real time digital simulator (RTDS) to simulate real-world events. The event cause analysis is performed using only one cycle of the voltage waveforms after the event is detected. The simulation results show the effectiveness of the proposed machine learning-based method compared to the state-of-the-art classifiers.**

## I. INTRODUCTION

Sensors and measurement devices have been evolving during recent years with different applications in power systems. The evolution has started with simple devices such as chart recorders and analog meters to more sophisticated devices such as phasor measurement units (PMU) [1]. These devices are capable of sending data up to one frame every 1 to 4 seconds (SCADA) or 30 to 120 frames-per-second (PMUs and μ-PMUs). While PMUs are very capable in steady-state and nominal frequency conditions, their measurements may have errors under transient conditions and non-nominal frequencies. However, in recent years, there has been more interest in using the next generation of high fidelity measurement devices, known as synchro waveform measurement units (SWMU) [1]-[2] or continuous point-on-wave (CPOW) units. These devices have the capability of measuring the time-synchronized voltage and current waveforms in high sampling rates and capture the transient behavior of power networks in a wide-area manner. They can be used for several applications in both power system operation and planning. In power system operation, they can be used for real-time wide-area situational awareness, state estimation, voltage stability monitoring, and protection coordination. In power system planning, they can be used for model validation, frequency response analysis, and event cause analysis. In this paper, we focus on the former application, and we use the terminology "SWMU". SWMUs installed along network lines or inside substations can be used for monitoring and analyzing system conditions when a disturbance occurs.

Event cause analysis has been studied in several previous works. Event classification using PMU dataset has been studied in [3]-[9]. The use of high sampling rate measurement devices for event cause analysis in transmission grids have previously been studied in [10]. However, with the increase in the complexity of distribution networks due to the integration of a vast number of renewable energy resources, and the lower voltage deviations in distribution grids, events detection and classification requires more advance measurement devices, such as the SWMU. In addition, there are some events that can only be identified with high sampling rates waveforms [11]. In this regard, this paper proposes an event cause analysis based on synchro waveform measurements. The proposed methodology is based on the convolutional neural network (CNN).

The use of CNN for event cause analysis in power networks is presented in a few works [12]-[15]. However, the proposed methods in these works are not comprehensive as they only use the time-synchronized measurement from one location or several unsynchronized measurements. In addition, some of them are not using synchro waveform measurements. In [14], CNN is incorporated for event cause analysis based on the PMUs data. The measurements from one substation are transformed into 2D inputs that result in temporal learning. In [15], the authors propose a fault detection and classification method using CNN along with the sparse autoencoder. The time-synchronized voltage and current measurements are captured from only one measurement device for temporal feature extraction and event cause analysis.

Iman Niazazari and Hanif Livani are with the Department of Electrical and Biomedical Engineering, University of Nevada, Reno, NV 89557 (niazazari@nevada.unr.edu).
Amir Ghasemkhani is with the Department of Computer Science and Engineering, California State University Sanbernardino, San Bernardino CA 92407.
Yunchuan Liu and Lei Yang are with the Department of Computer Science and Engineering, University of Nevada, Reno, NV, USA 89557.

This material is based upon work supported by the Department of Energy National Energy Technology Laboratory under Award Number DE-OE0000911. 

This paper presents a wide-area data-driven cause analysis of events in distribution power networks. The contributions of the paper are summarized as follows:

- From the methodology point of view, since the measurements have a close spatiotemporal correlation to each other, incorporating CNN for feature extraction results in capturing the spatiotemporal correlation of the wide-area synchro waveform measurements by convolving several filters through the stream of measurements. CNN takes advantage of the strong dependency existing among neighboring samples in measurements. The proposed spatiotemporal method results in better event cause analysis compared to existing only spatial or temporal feature learning methods. It outperforms other methods where they either did not take advantage of the spatiotemporal of measurements or they did not consider the possibility of simultaneous multiple high-frequency measurements. This is confirmed by comparing the results with other methods, autoencoder, support vector machine (SVM) and tapered multilayer perceptron (t-MLP) neural network.
- From the application point of view, important events including capacitor bank switching, transformer energization, fault, and high-impedance fault (HIF), with very similar spatial or temporal features in distribution networks are considered. As an example, identifying HIFs through protection devices are very difficult as they only have a small change in the fundamental component of the phasor measurements. Also, they pose a big safety concern for utilities as they usually occur when a conductor comes in touch with dry and very high impedance ground. Therefore, identifying them is vitally important from both operational and safety standpoints. As a result, cause analysis of such events is a very challenging task that requires a sophisticated classification algorithm. Moreover, the event cause analysis is performed using only one cycle of voltage measurements without other types of measurements such as current or frequency. Furthermore, the proposed method is validated via realistic synchro waveform data generated by the real-time digital simulator (RTDS).

The rest of this paper is organized as follows: Section II explains the event cause analysis framework. Section III presents the results and discussions. Finally, the conclusion and future works are presented in Section IV.

## II. Proposed Event Cause Analysis Framework

This paper proposes the use of the convolutional neural network (CNN) to learn the spatiotemporal features of synchro waveforms of voltage measurements. CNN is a class of deep neural networks and a powerful tool that can be used for spatiotemporal feature representation of data stream. Fig. 1 shows the basic CNN architecture. The convolutional layer makes use of a set of learnable filters to apply them to the input data. Filters are utilized to identify the presence of particular features in the input. Filters are slid across the width and height of the input signal, and a dot product is computed to result in an activation map. Different filters that detect different features are convolved on the input voltage data represented in time and location, and a set of activation maps is outputted which is passed to the next layer in the CNN. There may be some pooling layers interspersed between the convolutional layers to reduce the number of parameters and computation in the network, and also controlling overfitting by gradually reducing the spatial size of the network. Afterward, the extracted features go to the fully connected layer where the neurons have a complete connection to all the activations from the previous layers. Finally, for the classification of the classes, the extracted features go to the classification layer. In this paper, a softmax function is incorporated for the classification layer [5]. For further details on CNN see [16].

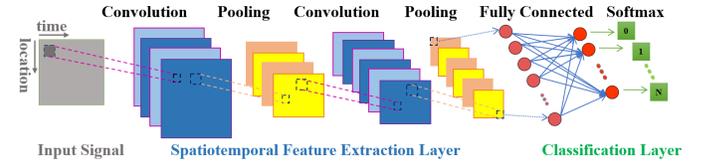

Fig. 1. Proposed method event classifier using CNN

The flowchart of proposed event cause analysis framework is shown in Fig. 2. The steps are summarized as follows:

(i) Synchronized waveforms of voltage measurements are measured at the designated buses and sent to the control center via available communication platforms such as a cellular network.
(ii) Modal transformation is applied to the measurements to calculate the mode-1 voltage for each designated bus.
(iii) Discrete wavelet transform (DWT) is applied to the mode-1 voltage for each designated bus to obtain the high and low-frequency contents in level-1 using the Daubechies-4 (db4) mother wavelet.
(iv) The absolute value of the wavelet transform coefficients (WTCs) is obtained and then normalized with respect to its peak for each designated bus ($|WTC|/|WTC|_{max}$).
(v) The normalized WTCs from each location are stacked on the top of each other forming the input matrix.
(vi) The input matrix is then applied to CNN for feature learning.
(vii) The extracted features are then used as the input to the softmax function for event cause analysis.

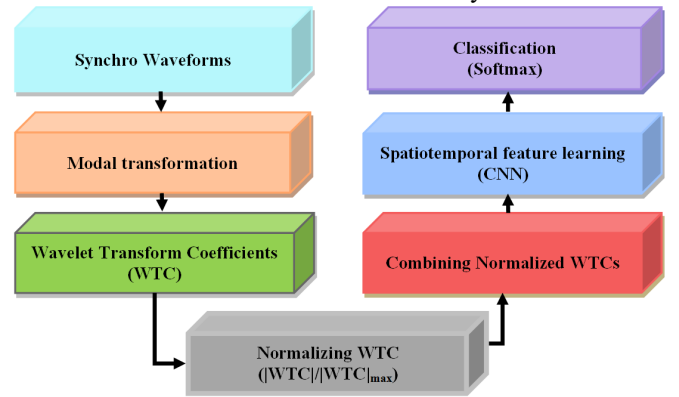

Fig. 2. Proposed event cause analysis flowchart based on synchro waveforms

## III. Results and Discussions

To validate the proposed methodology, the IEEE 13-bus test system is simulated using the RTDS, as a sample distribution feeder. Fig. 3 shows the location of SWMUs in the network along with the location of the events. Three software-in-the-loop SWMUs are located at buses 632, 671, and 675 which measure the voltages in these buses. Four events are simulated in this paper as follows:

- *Class 1 (Capacitor Bank Switching):* Switching of capacitor banks create high-frequency transients due to the inrush current coming from the system sources [17]. This event is simulated for different inception angles of switching and different values of the capacitor bank located at bus 675. The total number of instances of this class is 64.
- *Class 2 (Transformer Energization):* transformer energization can happen when a load is connected or when the transformer is disconnected and reconnected due to equipment degradation or failure. This event is simulated for different inception angles of the voltage waveform at different values of the transformer tap changer located between buses 633 and 634. The total number of simulated instances of this class is 144.
- *Class 3 (Fault):* All types of faults including LG, LL, LLG, and LLLG are simulated at four buses of 632, 634, 675, and 680. In generating the fault datasets, different fault characteristics including types, fault resistance, fault inception angles, and locations are considered. The total number of fault class instances is 320.
- *Class 4 (High Impedance Fault (HIF)):* HIFs usually occur because of unwanted electrical contact between a conductor and high impedance trees, or between a broken conductor and the ground. It can be unsafe due to fire hazards due to arching or touching the energized conductors [18]. Since the fault current remains below the threshold of overcurrent relays, detecting this type of fault is a much more challenging task. The total number of HIF instances is 72.

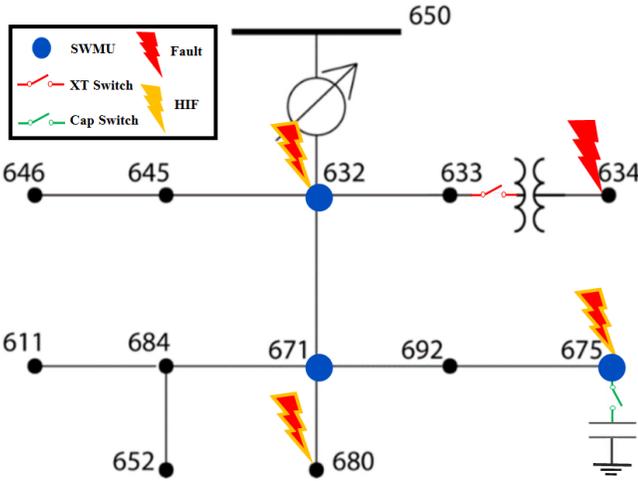

Fig. 3. IEEE 13-bus test system with events

Fig. 4 shows the RTDS implementation of the IEEE 13-bus test system and its interface to the event cause analysis platform. The dataset created by the RTDS is fed to the CNN based platform for the event cause analysis. All the event analysis has been carried out using MATLAB R2018a. In addition, the experimental platform used is Win 7; Intel Core i7-6700 with CPU 3.40 GHz and RAM 8 GB.

The initial weights for all the CNN layers are set with a normal distribution with zero mean and a standard deviation of 0.01, and the initial value of biases are zero. Numerous numbers of training and evaluation scenarios are carried out to find the best CNN parameters, i.e., the number of filters, convolutional layers, fully connected (FC) layers, and size of the filters. The optimal CNN structure is as follows: one convolutional layer with ten 2×20 filters, the stride of 1×1, maximum pooling of 1×2, and one fully connected layer. The classifier is trained on a single CPU using the stochastic gradient descent method with momentum with the number epoch of 50, mini-batch size of 8, the initial learning rate of 0.0001, and momentum of 0.9. Therefore, in all the following discussions, we use these selected CNN parameters. In addition, for the evaluation, 80% of the data set is used for training, and the rest is used for testing. In the following discussion parts, the effect of sampling rate, number and location of the measurement devices on the accuracy are studied. Finally, the performance of the proposed CNN-based method is compared with other methods.

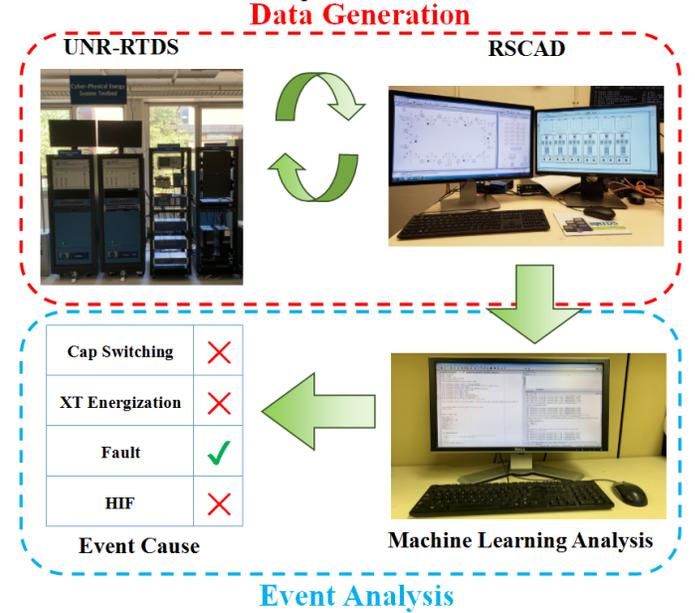

Fig. 4. Machine learning-based event cause analysis framework

*A. Sampling Rate Analysis*

One important aspect of event cause analysis with SWMUs is the sampling rate. Sampling rates of SWMUs can range anywhere between a few kHz to a few MHz and their prices change as the sampling rate increases. Therefore, performing a sensitivity analysis on the sampling rate of SWMUs and observing its effect on the event cause analysis results help the utilities to decide wisely on what product they should purchase.

Fig. 5 shows the event cause analysis with respect to the sampling rates. The sampling rates of SWMU start from 1.25 kHz and increases up to 20 kHz. As the RTDS simulation time step is bounded to 50 μs, we are not able to exceed the 20 kHz sampling rate which is the rate of sampling in many commercially available devices. As can be seen, as the sampling rate increases, the classification accuracy increases, too. The accuracy goes up from 67.5% for the 1.25 kHz sampling rate to 95% for the 20 kHz sampling rate, consequently, make evident a higher sampling rate results in better system performance. Therefore, the sampling rate of 20 kHz is recommended, and we use this sampling rate to perform the remaining analysis.

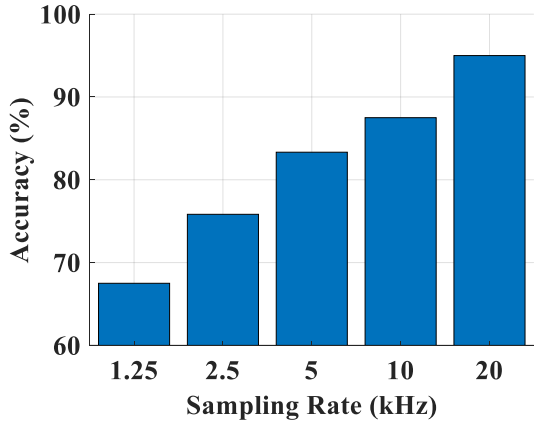

Fig. 5. Event cause analysis performance with respect to the sampling rates

### B. Impact of Number and Location of Measurement Units

SWMUs are among the most expensive advanced monitoring devices that can be installed in the power grid. In this regard, the knowledge on the performance of the event cause analysis for the different number of units and their locations is important for utilities to decide on the installation of these expensive measurement devices at certain locations. To assess the impact of the number and locations of SWMUs on the overall accuracy of event cause analysis, the following case studies are considered: Three cases with one units (at bus 675, 671, or 632), three cases with two units (at buses 632 & 671, 671 & 675, or 632 & 675), and one case with three units (at buses 632 & 671 & 675). The sampling rate for all cases is 20 kHz. Fig. 6 shows the accuracy in each case. As can be seen, generally as the number of measurement units is increased, the accuracy of the cause analysis is enhanced. This is due to the fact that an increase in the number of installed SWMUs leads to transmitting more high sampling data from different locations of the network, which consequently improves the feature learning and cause analysis performance. It can be observed that using two SWMUs at buses 671 & 675 or 632 & 675, we achieve almost the same accuracy as when all the SWMUs are used (94.17% ≈ 95%). Therefore, installing only two SWMUs can save money while delivering the same level of performance. Furthermore, it can be seen that by installing one SWMU at bus 632, achieves the accuracy of above 90%. Therefore, depending on the level of desired performance, installing only one SWMU at bus 632 might be sufficient.

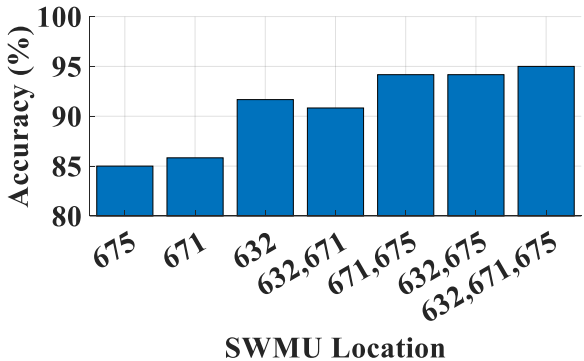

Fig. 6. Event cause analysis accuracy with respect to the number and location of SWMU

### C. Comparison with State-of-the-Art Methods

To further validate the effectiveness of the proposed spatiotemporal method, we compare its performance with the energy-based methods, such as autoencoder and support vector machine (SVM) where the features are obtained for a fixed number of intervals in time or space using mathematical operations, such as mean, summation, Euclidian norm, infinite norm and then feed to the classification process [7]. In addition, the results are compared with non-spatiotemporal methods, such as tapered multilayer perceptron (t-MLP) neural network [19].

To have a better insight into the classifiers' performance, four additional classification metrics other than accuracy (ACC) are calculated as well. These metrics are precision (PRE), recall (REC), F1 score (F1), and false positive rate (FPR). These metrics are defined for a binary classification problem as follows:

$$PRE = TP/(TP + FP) \quad (1)$$
$$REC = TP/(TP + FN) \quad (2)$$
$$F_1 = 2 \times (PRE \times REC)/(PRE + REC) \quad (3)$$
$$FPR = FP/(FP + TN) \quad (4)$$

where TP is True Positive, which is the number of events that are correctly predicted to fall into the target class; FP is False Positive, which is the number of events that are incorrectly predicted to fall into the target class; FN is False Negative, which is the number of events that are incorrectly predicted to fall out of the target class; and TN is True Negative, which is the number of events that are correctly predicted to fall out of the target class.

Table I shows the results of four different classification methods. As can be seen, the proposed spatiotemporal feature learning method outperforms the other methods. These results validate the superiority of the proposed spatiotemporal-based feature learning that extracts the spatiotemporal correlation of the stream of voltage measurements from different locations. In contrast, the energy-based methods disregard the spatiotemporal information as the energy of the signals in several or the entire sampling interval is calculated and utilized.

In our multiclass problem, the metrics are still the same as the ones used in the binary classification. However, the metrics are calculated for each class by treating it as a binary classification problem after combining all non-target classes into the second class. Then, the binary metrics are averaged over all the classes to get either a macro average (treat each class equally) or a micro average (weighted by class frequency) metric [20].

TABLE I
EVENT CAUSE ANALYSIS ACCURACY FOR THE IEEE 13-BUS TEST SYSTEM USING ENERGY-BASED, NON-SPATIOTEMPORAL FEEDFORWARD AND CNN-BASED SPATIOTEMPORAL METHOD

| | Method | ACC | PRE | REC | F1 | FPR |
|---|---|---|---|---|---|---|
| Energy-based | Autoencoder | 62.50 | 40.26 | N/A | N/A | 11.49 |
| | SVM | 76.67 | 80.22 | 69.75 | 74.62 | 8.76 |
| Non-Spatiotemporal | t-MLP | 87.50 | 91.35 | 83.40 | 87.19 | 5.36 |
| Spatiotemporal | CNN | 95.00 | 93.36 | 94.87 | 94.11 | 1.86 |

Fig. 7 (a) and (b) show the confusion matrix using the autoencoder and CNN methods where 80% of the data set is used for training and the remaining ones are used for evaluation.

The confusion matrix shows the performance of the event cause analysis method for distinguishing the correct events versus the misidentified ones. The rows show the predicted class (Output Class) and the columns correspond to the actual class (Target Class). The diagonal cells correspond to events that are correctly predicted, and the off-diagonal cells correspond to events that are incorrectly predicted. Both the number of events and the percentage of the total number of events in each case are shown in each cell. The column on the far right of the confusion plot displays the precision and its error for each individual class. The row at the bottom of the plot displays recall and its error. Finally, the cell in the bottom right of the plot shows the overall accuracy and the overall error. It can be seen that even though the autoencoder method is successful in distinguishing class 3, and somehow class 2, it completely fails to distinguish class 1 (capacitor bank switching) and class 4 (HIF) from other classes with 0% precisions. However, our proposed method significantly outperforms the non-convolutional method with 100% and 80% precisions in distinguishing classes 1 and 4, respectively. Furthermore, it can be seen that the overall accuracy in the CNN outperforms the t-MLP with 95% against 67.5%.

Fig. 7. Confusion matrix using (a) autoencoder and (b) CNN methods in the IEEE 13-bus test system

## IV. CONCLUSION

This paper presents a convolutional neural-network-based method for via event cause analysis in distribution networks. The results of this work can benefit the utilities to increase the events root cause analysis, enhance end-of-the-year power quality assessment, and abnormal events cause analysis. The method is developed based on data captured by a new type of measurement device, synchro waveform measurement unit (SWMU) or continuous point-on-wave (CPOW) units. The proposed method is used to detect four types of events namely, capacitor bank switching, transformer energization, fault, and high-impedance fault (HIF). As these events may not be easily identified by simply monitoring the relay voltage and current outputs or the peak values or duration of the time-frequency domain components of high-frequency voltage or current measurements. The proposed framework can be used in to increase the situational awareness of events for system operators. The results show the satisfactory performance of the proposed approach in comparison with other state-of-the-art methods. As future works, the authors will implement the proposed event cause analysis framework on larger networks with more events with real-world data sets. In addition, other novel classification methods will be explored.

5